\documentclass{aastex}
\usepackage{emulateapj5}
\begin{document}
\title{Constraining the magnetic effects on H\,{\sc i} rotation curves
and the need for dark halos}
\author{F.J. S\'{a}nchez-Salcedo }
\affil{Instituto de Astronom\'\i a, UNAM, Ciudad Universitaria, Aptdo. 70 264,
C.P. 04510, Mexico City, Mexico}
\email{jsanchez@astroscu.unam.mx}
\and
\author{M. Reyes-Ruiz}
\affil{Instituto de Astronom\'\i a, UNAM, Aptdo. 877, Ensenada, B.C. 22800,
Mexico}

\begin{abstract}
The density profiles of dark halos are usually inferred from the rotation
curves of disk galaxies based on the assumption that the gas is a good
tracer of the gravitational potential of the galaxies. Some
authors have suggested that magnetic pinching forces could alter 
significantly the rotation curves of spiral galaxies.
In contrast to other studies which have concentrated in the 
vertical structure of the
disk, here we focus on the problem of magnetic confinement in the
radial direction to bound the magnetic effects on the H\,{\sc i}
rotation curves. It is shown that azimuthal magnetic fields can
hardly speed up the H\,{\sc i} disk of galaxies as a whole. In fact, 
based on virial constraints we show
that the contribution of galactic magnetic fields to the 
rotation curves cannot be larger than $\sim 10$ km s$^{-1}$ at the
outermost point of H\,{\sc i} detection, if the
galaxies did not contain dark matter at all, and up to $20$ km s$^{-1}$
in the conventional dark halo scenario.
The procedure to estimate the maximum effect of magnetic fields
is general and applicable to any particular galaxy disk.
The inclusion of the surface terms, namely the
intergalactic (thermal, magnetic or ram)
pressure, does not change our conclusions.
Other problems related with the magnetic alternative to dark halos are
highlighted.
The relevance of magnetic fields in the cuspy problem of dark halos
is also discussed. 

\end{abstract}
\section{Introduction}
The rotation curves of spiral galaxies, especially dwarf galaxies
and low surface brightness galaxies, are used
to derive the profiles of the dark halos which are very useful as
tests for the standard cold dark matter scenario (e.g., de Blok
\& Bosma 2002). The
usual assumption is that the dynamics of the
neutral atomic gas is a good tracer of the radial gravitational force.
It can be shown that 
although spiral waves and non-circular motions may contribute to produce
substructure in the rotation curves (e.g., de Blok, Bosma \&
McGaugh 2003), they do not alter the rotation curves
significantly for the galaxies selected in
these investigations.
Galactic magnetic fields could also affect the gas dynamics 
in spiral galaxies \citep{pid64,per86,bat92,san97,ben00,bec02,bec03}.

The random small-scale component of the galactic magnetic fields acts as a 
pressure, giving support to the disk and, therefore, leading to 
a rotation (slightly) slower than the gravitational circular speed.
By contrast, some configurations of the large-scale magnetic field can
give rise to a faster circular velocity.
\citet{nel88} and \citet{bat92} proposed the so-called
``magnetic alternative'' to dark matter in which the contribution
of the magnetic pinch could be large enough to
explain the observed rotation curves of spiral galaxies without the necessity
of dark matter. This is an interesting possibility because some
universal properties such as the disk-halo conspiracy or the 
H\,{\sc i}-dark matter
constant relation, first noticed by \citet{bos78},
would have a natural explanation in this scenario
\citep{sn96a}, as well as the truncation of stellar
disks \citep{bat02}. However, after the publication
of the paper by \citet{bat92}, a plethora of possible difficulties 
faced by the model was pointed out by different authors
\citep{cud93,jok93,per93,val94,kat94,pfe94,sn96b,san97}.
In an extensive review,
\citet{bat00} argue that all
the shortcomings can be overcome and, therefore, magnetism
can explain the way in which galaxies rotate, without the help
of dark matter.

A certain MHD configuration has the ability of producing a faster
rotation of the whole plasma in the system, i.e.~magnetic confinement,
when the net contribution of the magnetic fields to the virial
theorem is negative (in the conventional notation).
In an isolated system, it is well known that the magnetic field contribution
to the virial theorem, applied to a large enough volume,
is {\it positive or zero}, but never negative, 
reflecting the net expansive tendency of magnetic fields
(e.g., Shafranov 1966). \citet{jok93} used this argument to
conclude that in the presence of magnetic fields, galaxies would need
even more dark matter than in the unmagnetized case.
One could argue that, since the virial theorem uses
global (volume-integrated) variables, it is possible that magnetism
has a ``centrifugal'' action in the vertical direction
but centripetal in the radial direction (radial confinement) and,
therefore, the expansive tendency of magnetic fields manifests only
in the vertical direction. We advance that this claim is very
wrong. An alternative way out is to assume that galaxies are not at 
isolation but embedded in the ubiquitous (magnetized)
intergalactic medium which would be responsible
for the required radial confinement.

Motivated by these ideas and given the importance of a correct 
interpretation of the rotation curves for Cosmology, 
we will try to put bounds on the
effects of magnetic fields on the H\,{\sc i} rotation curves.
However, we do not restrict ourselves to discuss the viability of
explaining rotation curves without dark halos; the orbits of Galactic
globular clusters, the satellite galaxies and possibly the dynamics of
remote stars in the outer Galactic halo, which are not affected by magnetic
fields, strongly argue in favor of a dark halo around the Milky Way
(e.g.~Freeman 1997). 
The main issue is to see whether galactic magnetic
fields can alter appreciably the H\,{\sc i} rotation curves, especially
in external galaxies for which the dark matter content is almost entirely
derived from the rotation curve of the gas.
Only if the answer is negative, we can be confident that the profiles
of dark halos, the dark matter content in galaxies as well as
other interesting empirical correlations
found between, for instance, halo parameters and luminosity are reliable.

Our investigation will be focus on the {\it radial} equilibrium
configuration of magnetized rotating disks. For reasons that should 
become clear later, the analysis of the radial equation of motion 
puts stringent bounds
on the maximum magnetic effects to the rotation curves; this being 
free of other complicating assumptions that appear in studies of
the vertical equilibrium configuration such as the external pressure,
the vertical dependence of the magnetic field within the height
of the H\,{\sc i} thin disk, or the flattening of the dark halo, which are
all very uncertain.  

The paper is organized as follows. In \S \ref{sec:ingredients} we
describe the ingredients of a simplified model in which magnetic
fields could give rise to a faster rotation in the outer parts
of H\,{\sc i} disks. In \S \ref{sec:core} the maximum effect on
the rotation curves is studied. Other difficulties inherent
to the magnetic alternative are discussed in \S \ref{sec:caveats}.
The case of magnetic fields producing a slower rotation and their
role as a possible remedy to the cuspy problem
of dark halos is left to \S \ref{sec:dwarf} and the final remarks are given 
in \S \ref{sec:final}.

\section{Ingredients of the Magnetic Model and Basic Theory}
\label{sec:ingredients}

In this Section, we describe some basic concepts useful for
understanding the remainder of the paper. The issue of the problem
of radial confinement will be addressed in \S \ref{sec:core}. 

\subsection{The One-dimensional Case}
\label{sec:oned}

In presenting the role of the magnetic tension in the H\,{\sc i}
rotation curves, the equations are simplified by assuming 
an equilibrium, steady-state, axisymmetric disk of a slightly ionized
gas with a purely azimuthal large-scale magnetic field $B_{\phi}$
(e.g., Battaner et al.~1992; S\'anchez-Salcedo 1997). 
The radial component of the equation of motion of the gas at $z=0$ 
in cylindrical coordinates reads
\begin{equation}
v_{\phi}^{2}(R)=R\frac{d\Phi}{dR}+\frac{R}{\rho}\frac{dP}{dR}+
\frac{1}{8\pi R \rho} \frac{d(R^{2}B_{\phi}^{2})}{dR},
\label{eq:vphi}
\end{equation}
where $\rho (R)$ is the gas density at $z=0$, 
$v_{\phi} (R)$ the circular speed of the gas,
$\Phi (R)$ the gravitational potential created by all masses
in the system, and $P(R)$ is the
gas pressure consisting of the thermal, turbulent, convective, and cosmic
ray components
plus also the magnetic pressure arising from the random magnetic
field component, but does not include the magnetic pressure arising
from the large-scale magnetic field.
All these variables should be taken at the midplane of the disk.
In equation (\ref{eq:vphi}) it was assumed $v_\phi \gg v_R, v_z$,
and thus we kept only the dominant terms associated with $v_\phi$.
In terms of the total pressure $P_{\rm tot}(R)=P(R)+B_{\phi}^{2}(R)/8\pi$,
it can be rewritten as:
\begin{equation}
v_{\phi}^{2}(R)=R\frac{d\Phi}{dR}+\frac{R}{\rho}\frac{dP_{\rm tot}}{dR}+
\frac{B_{\phi}^{2}}{4\pi\rho}.
\label{eq:vphipres}
\end{equation}
The second term of the RHS is the radial pressure force that
acts inwards or outwards depending on 
whether $P_{\rm tot}$ increases outwards or inwards.
The third term is the magnetic tensile force which always imposes a net
inward force.

We should remind that ordered galactic-scale
magnetic fields are not expected to be a general feature
in all disk galaxies, especially in dwarf galaxies. 
Although observations of magnetic fields in
these galaxies are scarce, they do not indicate magnetic fields
with a global symmetry, but a
much more chaotic galactic-scale field than in normal spirals (e.g.,
Chyzy et al.~2000). Therefore, models based on a regular azimuthal magnetic
field may not be an adequate representation for all of them.
In fact, indirect observations suggest that regular magnetic fields may
be ignorable, at least in some dwarf galaxies. The dominance
of spherical holes in the H\,{\sc i} of the nearby dwarf galaxy
IC 2574 has lead \citet{wal99} to conclude that
large-scale magnetic fields are dynamically unimportant in
the outer parts of this galaxy.

As it is shown in Appendix A, $P(R)$ is always a decreasing
function of $R$ in approximately-isothermal, exponential galactic disks, 
even if the disk were embedded
in an ambient medium with high pressure. Therefore, the asymmetric
drift always produces a force pointing outwards
and hence it can never contribute to enhance the circular speed of the gas.
Moreover, the asymmetric drift is almost negligible in rotating
galactic disks except in
very low-mass galaxies such as DDO 50 \citep{car90,deb02,bur02}. 
Thus we will also ignore it, hence $dP_{\rm tot}/dR\approx
(1/8\pi)dB_{\phi}^{2}/dR$, and the force balance equation is
simplified to
\begin{equation}
v_{\phi}^{2}(R)=v_{\rm c}^{2}(R)+
\frac{1}{8\pi R \rho} \frac{d(R^{2}B_{\phi}^{2})}{dR},
\label{eq:vphisecond}
\end{equation}
where $v_{\rm c}^{2}\equiv Rd\Phi/dR$ is the gravitational circular
velocity.
Note that the magnetic pressure arising from the large-scale field
is kept in our analysis because it cannot be neglected in general. The role
of the magnetic pressure term associated with the random (small-scale)
magnetic field component is discussed in detail in \S \ref{sec:dwarf}.
It should be also clear that the vertical gradient of $P$ is not
negligible when considering the vertical configuration of the disk,
since it is the main agent responsible for the vertical support 
against gravity.

From Eq.~(\ref{eq:vphisecond}) we see that for a purely azimuthal magnetic
field decaying radially not faster than $R^{-1}$,
the magnetic tension force produces radial confinement. 
Such configuration is usually
called a linear or Z-pinch. The resulting inward
magnetic force, the pinch, accounts for the required centripetal
force responsible for an anomalous faster rotation.
Our aim is to constrain the maximum
effect of the Z-pinch in galactic disks from very general grounds.
It is worth mentioning that the Z-pinch configuration 
is not used in tokamaks as interchange and other instabilities
quickly destroy the magnetic configuration (e.g., Roberts 1967).
Restricted to purely azimuthal fields,
the force-free magnetic configuration occurs for a field profile
$B_{\phi}=B_{0}(R_{0}/R)^{2}$, where $B_{0}$ is the magnetic
strength at some radius $R_{0}$. For that profile, the outward
magnetic pressure force is in balance with the inward tensile
force at any radius. This profile is, however, rather unphysical
because it is singular at $R=0$.

Observations of galactic magnetic fields suggest
that the radial component of the field, $B_{R}$, 
is comparable to $B_{\phi}$ at least in the optical
disk and hence neglecting $B_{R}$ may not be a good approximation. 
One could argue, however, that the magnetic pitch angle becomes
very small in the outermost parts of galactic disks, where 
measurements of the field orientation are scarce.
Since $B_{R}$ is not well-suited to produce radial confinement,
the most optimistic and simplest situation, in order
to maximize the magnetic effects on the rotation curves, occurs 
when the field is taken as purely azimuthal. 
As we are interested in putting upper limits on
the role of magnetic fields we will assume throughout this
paper that $B_{R}=0$ in the outer disk to give the model as
much leeway as possible, as it is assumed in the
magnetic alternative.

An important contribution to $P$ is the turbulent pressure.
For a certain component of the interstellar medium, the averaged
turbulent pressure is given by
$\rho v_{\rm t}^{2}/2$, where $v_{\rm t}$ is the 
one-dimensional turbulent velocity dispersion. For the H\,{\sc i}
components, $v_{\rm t}\approx 6$--$10$ km s$^{-1}$ (e.g., Boulares
\& Cox 1990) and is nearly constant with galactocentric radius and
also from galaxy to galaxy (Sellwood \& Balbus 1999, and
references therein). The magnetic field is said
to be in equipartition with the turbulent motions when the gas-to-magnetic 
field pressure ratio is $1$, or equivalently, when the turbulent velocity
dispersion coincides with the Alfv\`en speed. In the literature, one often 
finds the statement that dynamo action in a thin disk saturates at 
equipartition field strengths (e.g., Ruzmaikin, Sokoloff \& Shukurov 1988;
Shu 1992; Poezd, Shukurov \& Sokoloff 1993; Beck et al.~1996; Kulsrud 1999).
Therefore, it is interesting
to estimate those values of $v_{\phi}-v_{\rm c}$ for which the 
large-scale magnetic field would be in superequipartition; this is done 
in the next. Denoting by $v_{\rm A}$
the Alfv\`{e}n velocity associated with the regular field,
we can write
$v_{\phi}^{2}-v_{\rm c}^{2}\equiv(8\pi R \rho)^{-1} d(R^{2}B_{\phi}^{2})/dR=
v_{\rm A}^{2}+(R/8\pi\rho)dB^{2}_{\phi}/dR$.
In the likely case that the strength of the large-scale magnetic field does not
increase with galactocentric radius it implies
$v_{\phi}^{2}-v_{\rm c}^{2}\lesssim v_{\rm A}^{2}$.
Hence, for changes in the rotation velocity of
$v_{\phi}^{2}-v_{\rm c}^{2}> v_{\rm t}^{2}$, 
the regular magnetic field should be above 
equipartition with the turbulent motions.
The above condition is equivalent to $v_{\phi}>\sqrt{v_{\rm t}^{2}+
v_{\rm c}^{2}}$ or, in terms of 
$v_{\phi}-v_{\rm c}$:
\begin{equation}
v_{\phi}-v_{\rm c}>(v_{\rm t}^{2}+v_{\rm c}^{2})^{1/2}-v_{\rm c}
\approx v_{\rm c}\left(1+\frac{1}{2}\frac{v_{\rm t}^{2}}{v_{\rm c}^{2}}\right)
-v_{\rm c}=\frac{1}{2}\frac{v_{\rm t}^{2}}{v_{\rm c}}.
\end{equation} 
For $v_{\rm c}\approx 100$ km s$^{-1}$, variations of 
$v_{\phi}-v_{\rm c}\gtrsim 0.5$ km s$^{-1}$ already require magnetic
fields over equipartition.
Such a situation is not very appealing to most dynamo theorists.
However, in order to present our results as generic as possible
and since there are some observations that could
suggest magnetic energy densities above the turbulent ones 
in the disk of galaxies (e.g., Beck 2002, 2003) and in the Galactic
center (e.g., Moss \& Shukurov 2001), we will not employ equipartition
when discussing the magnetic effects on the rotation curves. 

So far, the usual way to proceed was to integrate Eq.~(\ref{eq:vphi})
out to a certain radius
in order to derive the required radial profile for $B_{\phi}(R)$ to explain
the observed rotation curve $v_{\phi}$, given the
volume density at $z=0$ and  
the gravitational circular velocity (including a
contribution of the dark halo as in S\'anchez-Salcedo (1997) or 
only the luminous components as in Battaner et al.~(1992)).
The question that arises is whether these local solutions 
can be matched with boundary conditions at infinity or not.
This issue is very delicate and it will be proven very useful
to put bounds on the influence of galactic magnetic fields in the
gas dynamics at large scales.
Before that, however, it is illustrative to describe how galaxies
can reach a configuration with a faster rotation in order to
have a complete picture of the underlying physics. This is briefly
described in the next subsection.

\subsection{The Phase of Magnetic Contraction}
\label{sec:mcp}

The assumption that the magnetic field is
mainly azimuthal immediately suggests that it commenced to
grow after the galactic disk has formed. 
In the absence of magnetic tension, 
the {\it initial} galactic disk or protodisk
should have a rotation speed given by the mass distribution
$v_{\phi}^{2}=v_{\rm c}^{2}=R d\Phi/dR$.
If the rotation curves are significantly affected
by the magnetic field, one should expect that, as the magnetic field
is amplified no matter by what process, the disk initially in gravitational
force balance, will experience a radial contraction caused by
the growing magnetic tension, leading to a disk
with a faster rotation due to angular momentum conservation.
We will refer to this process as the magnetic contraction phase (MCP).
If the mass loss by outflows is unimportant,
the MCP is the only way capable of spinning up the disk
because the total angular momentum stored in the fields
($\propto \int {\mbox{\boldmath{$r$}}}\times [{\mbox{\boldmath{$E$}}}
\times {\mbox{\boldmath{$B$}}}]\,dV$) is negligible in both the initial
protodisk and in the final configuration. 
Other processes, such as turbulent viscosity, 
are very inefficient to produce the desired effect.

The conservation of angular momentum implies that, for a galaxy
similar to the Milky Way
with a dark halo and a rotation speed of $220$ km s$^{-1}$ at $30$ kpc,
an increase of $10$ km s$^{-1}$ in the velocity speed requires a
radial contraction of $1.3$ kpc at that radius. The contraction must be huge
in the magnetic alternative; 
see \S \ref{sec:truncation}.

\section{Galactic Disks with Faster Rotation. The Problem
of the Radial Confinement}
\label{sec:core}
\subsection{Isolated Galaxies}
\label{sec:isolatedgal}

In the previous section it was pointed out that there is a pinching
effect only if $B_{\phi}$ does not decrease faster than $1/R$.
However, this decay is untenable at large $R$ in real astrophysical systems
as it will be shown in the next
and, therefore, it is inevitable that beyond a certain radius, the
magnetic tension produces a force pointing outwards (see 
Fig.~1 for a visualization).
In fact, the slowest radial magnetic decay occurs in the ideal case
of having strict cylindrical geometry with currents, $J_{z}$, flowing
from $z=-\infty$ to $z=\infty$ (otherwise, if
currents are confined to a certain region, 
${\mbox{\boldmath $B$}}$ must decrease with radius 
at least as fast as a dipole field for large ${\mbox{\boldmath $r$}}$:
$B\leq K/r^{3}$, where $K$ is a constant). 
Under strict cylindrical geometry 
\footnote{In such a situation,
the magnetic field is always purely azimuthal, consistent 
with the assumptions given in \S\ref{sec:ingredients} and, hence, 
the essential physics of the 
mechanism giving rise to greater azimuthal velocities of the
H\,{\sc i} disk, is entirely contained within the 
one-dimensional radial models.},
the condition of closed current loops
is equivalent to imposing $\int_{0}^{R} J_{z} R'\,dR'\rightarrow 0$ 
at large $R$ (hereafter, this will be our definition of
{\it isolated system} in a cylindrical system). 
From Amp\`ere's law, this requirement implies that 
$B_{\phi}< C/R$ at large $R$, being $C$ a constant. 
So there must exist a certain radius denoted by $R_{\rm kep}$ at which 
the magnetic field starts to decay as $1/R$ or faster.
$R_{\rm kep}$ is therefore the radius at which the rotation
speed of the gas coincides with the local Keplerian velocity.
If $R_{\rm HI}$ is the radius of the last point of
H\,{\sc i} detection, we require $R_{\rm kep}>R_{\rm HI}$
in order to have an anomalous faster rotation along the observed disk.

The system will be in a configuration of equilibrium at $R>R_{\rm kep}$
only if the inward gravitational force is higher than the outward 
magnetic force 
\begin{equation}
R\frac{d\Phi}{dR}>-\frac{1}{8\pi R\rho}\frac{d(R^{2}B_{\phi}^{2})}{dR}.
\label{eq:directbal}
\end{equation}
If this condition is not fulfilled the disk will expand radially due to the
magnetic forces. This condition can be written as
\begin{equation}
B_{\phi}(R)>\frac{1}{R}\left[R_{\rm kep}^{2}B_{\phi,{\rm kep}}^{2}-
8\pi \int_{R_{\rm kep}}^{R} \rho\, R\, v_{\rm c}^{2}\,dR\right]^{1/2},
\label{eq:atrest}
\end{equation}
where $B_{\phi,{\rm kep}}$ is the magnetic field strength at  
$R_{\rm kep}$. Since a dependence $B_{\phi}\approx C/R$ at large $R$
is unphysical (see above), an upper limit to $B_{\phi,{\rm kep}}$ can be
established from Eq.~(\ref{eq:atrest}):
\begin{equation}
B_{\phi,{\rm kep}}^{2}\leq \frac{8\pi}
{R_{\rm kep}^{2}}
\int_{R_{\rm kep}}^{R_{B_{\phi}}} \rho(R,0)\,R\, v_{\rm c}^{2}(R)\,dR,
\label{eq:upperlimit}
\end{equation}
where $R_{B_{\phi}}$ is the radius where $B_{\phi}\approx 0$. In practice
we will take $R_{B_{\phi}}\rightarrow \infty$ to estimate the maximum
contribution. The maximum value for $B_{{\phi},{\rm kep}}$ 
corresponds to a situation in which there is balance at $R>R_{\rm kep}$
between gravitational and magnetic forces and thus 
the material outside the $R_{\rm kep}$ circle has no rotation
at all.

The upper limit derived above has an easy interpretation.
In principle, one could achieve very large differences 
$v_{\phi}-v_{\rm c}$ if one would make the density arbitrarily
small than the Alfv\`{e}n speed would be arbitrarily large.
However, according to Eq.~(\ref{eq:directbal}), this is not possible
globally because the magnetic field has to be held in by the radial
weight of the plasma in the radial gravitational field 
since the net effect of magnetic fields is never confining. 

It is customary and natural to discuss this 
kind of constraints using the virial theorem which says that the net 
nature of magnetic fields 
is expansive or null (but never confining)
in both the radial and vertical directions (Appendix B).
More specifically, for toroidal magnetic fields and strict cylindrical
geometry, the total rotational energy of an isolated system should
be the same with or without magnetic fields,
keeping the mass distribution fixed (so that $\Phi(R)$; see Appendix B
for further details). By imposing
that constraint, a similar upper limit for $B_{\rm kep}$ is found
(Appendix C), except by a factor ($\lambda_{\rho B}$), 
between $1$--$2$, that takes into account possible
three-dimensional effects.

It is convenient to consider first the extreme and hypothetical case that
spiral galaxies do not host dark matter halos. For a galaxy
with a mass in gas and stars $M$, $\Phi(R)\approx -GM/R$
in the outer parts and 
\begin{equation}
B_{\phi,{\rm kep}}^{2}\leq B^{2}_{\rm cr}\equiv 8\pi
\lambda_{\rho B}\frac{GM}{R_{\rm kep}^{2}}
\int_{R_{\rm kep}}^{\infty} \rho(R,0)\,dR.
\label{eq:upperkep}
\end{equation}
The assumption that $\Phi(R)\approx -GM/R$ is satisfactory for many high 
surface density galaxies
(e.g., van Albada et al.~1985) but it is not necessarily fulfilled
in some low surface brightness galaxies (LSB)
and dwarfs for which the gas surface density
is not exponential and its contribution to
the potential may be comparable or even higher than that from the stellar
component at radii $\sim R_{\rm HI}$ 
for certain values of the adopted stellar mass-to-luminosity ratio.
However, we are confident that our conclusions would not change 
for these galaxies even adopting the most unfavorable assumptions.

In the optimistic and rather unlikely situation that
$B_{\phi}$ were constant with $R$ out to $R_{\rm kep}$, 
i.e., $B_{\phi}(R)=B_{\rm cr}$ at $R<R_{\rm kep}$, the characteristic
variation $v_{\phi}^{2}-v_{\rm c}^{2}$ would be 
$v_{\phi}^{2}-v_{\rm c}^{2}\approx B_{\rm cr}^{2}/(4\pi \rho)$. 
By using Eq.~(\ref{eq:upperkep}), it is possible to estimate 
the maximum relative variation of the circular speed of
the gas at a radius $R<R_{\rm kep}$:
\begin{equation}
\frac{v_{\phi}^{2}-v_{\rm c}^{2}}{v_{\rm c}^2}\bigg|_{R}=2
\lambda_{\rho B} \left(\frac{R}{R_{\rm kep}}\right)
\left(\frac{R_{\rm g}}{R_{\rm kep}}\right)
\exp\left(\frac{R-R_{\rm kep}}{R_{\rm g}}\right),
\label{eq:ten}
\end{equation} 
where we have adopted a radial exponential decay for $\rho (R,0)$ with
scale $R_{\rm g}$ (e.g., Bland-Hawthorn, Freeman \& Quinn 1997).
Note that $(v_{\phi}^{2}-v_{\rm c}^{2})/v_{\rm c}^{2}$
is not sensitive to $P_{\rm ext}$ 
because even though the density at the midplane increases with $P_{\rm ext}$,
equation (\ref{eq:ten}) does not depend on the absolute value of the
density but only on $R_{\rm g}$.
The main source of uncertainty stems from the value of 
$R_{\rm kep}$, but the most optimistic situation occurs when
$R_{\rm kep}\approx R_{\rm HI}$, so we will assume for the discussion that 
both radii are identical, $\approx 30$ kpc for common spiral galaxies.
In order to estimate $v_{\phi}^{2}-v_{\rm c}^{2}$ from Eq.~(\ref{eq:ten}),
we will assume that the mass of gas in the disk beyond
$R_{\rm HI}$ and within $|z|<h_{1}$, being $h_1$ the 
characteristic scale height of the gas disk (see Appendix C for 
further details), is $0.5$ percent the total 
mass of the disk. This seems
to be the case of dwarf galaxies (van den Bosch et al.~2001), 
but it is probably a too generous value for normal spiral galaxies.
The latter assumption corresponds to a radial scale length for the
gas, $R_{\rm g}$, of $\sim 5.5$ kpc. Putting these values 
into Eq.~(\ref{eq:ten}), we obtain $(v_{\phi}^{2}-v_{\rm c}^{2})/v_{\rm c}^{2}=
0.36 \lambda_{\rho B}$ at $R=30$ kpc. If we take for
reference values $v_{\rm c}=100$ km s$^{-1}$ at $30$ kpc, 
the maximum change in the rotation velocity, 
$\Delta v\equiv v_{\phi}-v_{\rm c}$, ranges between $15$ to $30$ km
s$^{-1}$ for $\lambda_{\rho B}=1$ to $2$, respectively.

This value for $\Delta v$ should be regarded as a generous
estimate for two reasons. First, the azimuthal field is not
expected to be approximately
constant while the volume density of gas at $z=0$ varies by 
a factor of $\sim 50$--$100$ \citep{dip91}. 
Secondly, the adopted equilibrium configuration, namely 
$v_{\phi}(R)=0$ at any $R>R_{\rm kep}$, cannot be reached in practice,
starting from a rotating disk in equilibrium with the gravitational
force, due to the conservation of angular momentum. In fact, 
the magnetic field begins to grow in the disk
as discussed in \S \ref{sec:mcp}, creating a magnetic force directed outward
beyond $R_{\rm kep}$. This
region will undergo a continuous outward drift, while the inner
disk ($R<R_{\rm kep}$) must be subjected to the required contraction
phase, generating a gap in the surface density of the disk.
In order to quantify this effect, we assume for simplicity that the
outer disk ($R>R_{\rm kep}$) suffers an approximately self-similar
expansion with factor $\psi>1$, i.e., $\rho(R)=(\rho_{0}/\psi^{2})
\exp(-R/\psi R_{\rm g})$ at $R>\psi R_{\rm kep}$,
in the final equilibrium configuration. From the conservation of
angular momentum, it can be shown that the right-hand side of Eq.~(\ref{eq:ten})
is now a factor $\psi^{2}/(\psi-1)$ smaller (Appendix C). 
Hence, the optimized situation occurs for $\psi=2$ and then 
$(v_{\phi}^{2}-v_{\rm c}^{2})/v_{\rm c}^{2}<0.1 \lambda _{\rho B}$ 
at $R=30$ kpc.
This implies that $\Delta v_{\phi} < 10$ km s$^{-1}$ 
at the periphery of H\,{\sc i} disks in
normal spiral galaxies under the assumption that they 
contained only stars and the observed gas. Therefore, it is not
possible to banish dark halos by magnetic fields because one
would require much larger values of $\Delta v_{\phi}$.

For disk galaxies with a dark halo, as postulated in the conventional
scenario, it is possible to follow the same procedure to estimate
the maximum effect of magnetic fields when $v_{\rm c}=$cte,
instead of $v_{\rm c}^{2}=GM/R$ (see Appendix C). 
Similar arguments suggest that
the expected local, and probably transient, variations in the rotation speed 
larger than $\sim 20$ km s$^{-1}$ are very unlikely
even in models including dark halos (see also \S \ref{sec:mcp} and
S\'anchez-Salcedo 1997).

In the next subsections, it will be shown that these upper limits for
$\Delta v_{\phi}$ also apply even when galaxies are not
isolated systems but are embedded in an ambient medium, thereby making 
these estimates very robust.

\subsection{Non-Isolated Galaxies}
It could be argued that the model described by Eq.~(\ref{eq:vphi}) is
too simplistic because other terms not considered in that equation,
as a radial flow or other components of the magnetic field, 
might be important, regarding
the radial confinement, beyond $R_{\rm HI}$\footnote{Within the 
$R_{\rm HI}$ circle, observations suggest that other terms are unimportant
for those galaxies under consideration.}.
In fact, unlike the radial component of the field, the $z$-component may
contribute to the radial confinement of the disk and will be included
in our analysis. In addition, the hydrodynamical pressure
by a radial inflow due to intergalactic accretion of matter may also 
play a role. Since systems embedded in an ambient medium 
usually present sharp boundaries, it is
convenient to use an integral form of the equations
as that given by the virial theorem.

The scalar virial theorem for fields with $B_{R}=0$ and strict cylindrical
symmetry (i.e., $\partial/\partial z=\partial/\partial\phi=0$)
applied to a finite cylindrical volume $V$, bounded by a cylinder
$S$ of height $L$, can be written as:
\begin{equation}
2{\mathcal{T}}+2\int_{V} P\,dV-2P_{\rm S}V+
{\mathcal{W}}+{\mathcal{M}}_{\rm cyl}-\int_{S}\rho\,
({\mbox{\boldmath{$v$}}}\cdot
{\mbox{\boldmath{$r$}}})({\mbox{\boldmath{$v$}}}\cdot
d{\mbox{\boldmath{$S$}}})=\frac{1}{2}\ddot{I},
\label{eq:virialcyltxt}
\end{equation}
where ${\mathcal{T}}$ is the kinetic energy of the gas, ${\mathcal{W}}$
its gravitational energy in the gravitational field 
created by all masses in the system,
$P$ the gas pressure as defined in \S \ref{sec:oned}, $P_{\rm S}$ 
the gas pressure at the boundary surface, 
${\mbox{\boldmath{$r$}}}$ is a coordinate vector, $\ddot{I}\equiv
d^2 I/dt^2$ with $I$ the
trace of the moment of inertia tensor and ${\mathcal{M}}_{\rm cyl}$
is given by
\begin{equation}
{\mathcal{M}}_{\rm cyl}=\frac{1}{4\pi}\left(\int_{V} B_{z}^{2} \,dV-
\left(B_{z,{\rm S}}^{2}+B_{\phi,{\rm S}}^{2}\right) V\right),
\label{eq:magtermtxt}
\end{equation}
(e.g., Fiege \& Pudritz 2000 and Appendix B). 
Throughout this paper the subscript $S$
refers to quantities valuated at the radial surface of our
cylindrical volume with radius $R_{\rm S}$, and the subscript
`${\rm cyl}$' stands for the cylindrical case.
This form for the virial theorem is valid for any arbitrary helical
field ($B_{R}=0$) provided cylindrical geometry; no other additional 
assumption is required. The surface integral terms including the 
$z=\pm L/2$ surfaces are taken into account in Eq.~(\ref{eq:magtermtxt}).
The integration of the magnetic terms can be found in Appendix B.
For any further details in the derivation, 
we refer the reader to the paper by Fiege \& Pudritz (2000).
 
We can identify two new vias of radial confinement: 
by ram pressure [last term of LHS of Eq.~(\ref{eq:virialcyltxt})]
or by an external magnetic field.
The ambient pressure ($P_{\rm S}$), on the other hand, 
only produces vertical confinement as discussed in Appendix A.
If these new terms are able to produce radial confinement then the
constraint given by Eq.~(\ref{eq:upperlimit}) could be somehow relaxed. 
The possible role of these terms will be discussed in two separate sections.  

\subsubsection{Ram Pressure}
\label{sec:ramP}
So far, we have neglected the terms involving $v_{R}$, 
because H\,{\sc i} observations show
that $v_{R}$ is very small compared to $v_{\phi}$ for the galaxies
under consideration. Beyond the edge of the observed H\,{\sc i} disk,
the value of $v_{R}$ is uncertain.
We consider now 
the surface integral of the momentum stress in Eq.~(\ref{eq:virialcyltxt}),
which may have a negative sign for the case of strong mass accretion into
the disk. In fact, it can be shown that even though 
$\ddot{I}$ is expected to be negative, the combination
$-\ddot{I}/2 -\int_{S}\rho\,
({\mbox{\boldmath{$v$}}}\cdot
{\mbox{\boldmath{$r$}}})({\mbox{\boldmath{$v$}}}\cdot
d{\mbox{\boldmath{$S$}}})$
could be still negative, showing the confining effect of the ram pressure.
(For example, in the one-dimensional case of a supersonic flow  
that slows down to rest by crashing perpendicularly to a rigid plane 
wall and forms a strong adiabatic shock, $-\ddot{I}/2 -\int_{S}\rho\,
({\mbox{\boldmath{$v$}}}\cdot
{\mbox{\boldmath{$r$}}})({\mbox{\boldmath{$v$}}}\cdot
d{\mbox{\boldmath{$S$}}})=
-(2/3)\int_{S}\rho\,
({\mbox{\boldmath{$v$}}}\cdot
{\mbox{\boldmath{$r$}}})({\mbox{\boldmath{$v$}}}\cdot
d{\mbox{\boldmath{$S$}}})
$.)

Since spherical accretion cannot confine a disk, the
inflow should be extremely collimated parallel to the galactic plane,
i.e.~$v_{R}\gg v_{z}, v_{\phi}$, for which there is no
evidence. Moreover, this collimation is very unlikely to occur
if the disk forms from a sphere of gas. Gas cools and parcels of gas
acquire circular orbits, conserving approximately angular momentum,
when they collapse to settle into a disk,
leading to $v_{R}\rightarrow 0$ (see R\"{o}gnvaldsson 1999 for
results based on numerical simulations). Furthermore, in the hypothetical case
that such inflow were possible by some unknown mechanism, the accretion
speed should be so high that an accretion shock would be unavoidable,
transforming ram pressure into thermal pressure and probably
generating a hydraulic jump in the edge of the disk
and other observable features such as H$\alpha$ emission.
The hydraulic jump will divert the flow around the disk 
decreasing dramatically its confinement ability.

\subsubsection{Confinement by the Intergalactic Magnetic Field}
\label{sec:mag}

A coherent external (intergalactic) magnetic field could in 
principle contribute to the confinement under certain circunstances.
It is important to distinguish here between the intracluster and the
wider intergalactic field. As we are dealing with galaxies in the field,
we are interested in the amplitude of the latter one.
At present, setting observational constraints to the strength and
topology of the intergalactic magnetic fields is a challeging task
(see Ryu, Kang \& Biermann 1988;
Kronberg et al.~2001; and Widrow 2002 for a review) and, therefore,
they are very uncertain. Coherent magnetic fields on scales much
greater than the typical sizes of galaxies could be as large as
$\sim 0.1$ $\mu$G in the optimistic situation that extra amplification
through batteries, dynamos or other routes of magnetization were
operative.

We assume the ideal situation that initially the pregalactic 
magnetic field is uniform and fills
the protogalaxy making a finite angle $\alpha$ with the rotation
axis of the galaxy. 
As the disk is forming by accretion of gas, the magnetic field lines are
advected with the flow. It is well-known in studies of the collapse
of interstellar clouds that, under flux-freezing conditions, this
compression produces that the total
magnetic contribution to the virial theorem is positive (see, e.g.,
Chapt.~24 of Shu 1992 and Chapt.~11 of Mestel 1999) and, 
therefore, the confinement by the external field
is impeded. In order to have confinement, the disk must be able
to get rid of the magnetic flux.

In the strict cylindrical symmetry, the external magnetic field
must be vertical and then Eq.~(\ref{eq:magtermtxt})
with $B_{\phi {\rm S}}=0$ tells us that magnetic confinement is only
possible if the volume-averaged vertical component of the magnetic
field within $V$ is strictly smaller than the magnetic field at the boundary.
Since $v_{R}$ is directed inward and $\partial v_{R}/\partial R<0$
in the MCP, 
$|B_{z}|$ increases at every point of the disk due to the
inward advection (see also Reyes-Ruiz \& Stepinski 1996; Moss \& Shukurov 2001);
under those conditions confinement is not viable.
This problem could be alleviated if 
magnetic field lines were expelled from the disk by ambipolar and
turbulent diffusion. However, a thorough numerical study by \citet{how97}
shows that $|B_{z}|$ is also increased by
ambipolar shearing terms to a value considerably larger than its
initial value. 
In the lack of a mechanism able to reduce $|B_{z}|$ within disks, 
pregalactic magnetic fields do not help to confine galactic disks.
This example clearly illustrates why self-confinement is so
rare in nature.

Models based on the confinement
by anchored fields to the galactic nucleus will present
a similar problem, as $|B_{z}|$ is expected to
decline outwards.

\section{The Magnetic Alternative}
\label{sec:caveats}
Battaner \& Florido (1995, 2000) and Battaner, Florido \& Jim\'{e}nez-Vicente
(2002) pointed out that: (1) the magnetic alternative to dark halos
does not suffer from the problem of excessive flaring of the H\,{\sc i} disk
and (2) the existence of truncated stellar disks
is explained satisfactorily in the magnetic alternative.
Putting aside the problem of radial confinement discussed in
the previous sections,  
it is stimulating to see whether alternative scenarios to
conventional ideas have the potential of explaining other phenomena in
a natural way. If this occurs, alternative models could attrack much
more attention. For instance, some empirical correlations of galactic
warps have led to Castro-Rodr\'{\i}guez et al.~(2002) 
to suggest tentatively that outer rotation curves could not be caused by
the presence of a massive halo.
In the next sections we will show that the flaring
problem remains unsolved and that the existence of truncated stellar disks
cannot be considered a prediction of the magnetic alternative.

\subsection{The MCP and Implications for the Truncation of Stellar Disks}

\label{sec:truncation}
If galaxies contained only stars and the observed gas,
the initial protodisk of mass $M_{\rm i}$ should follow 
approximately a Keplerian rotation, $v_{\phi}^{2}=GM_{\rm i}/R$, at
large $R$, and evolves to the present-day 
exponential disk of mass $M_{\rm f}$ and radial scale length $R_{\rm d}$,
due to magnetic contraction.
The peak velocity of the final disk is 
$v_{\rm max}=0.62 \sqrt{GM_{\rm f}/R_{\rm d}}$
(e.g., Binney \& Tremaine 1987). For simplicity,
we assume that beyond $2.27 R_{\rm d}$ the rotation curve
of a present galaxy is flat with circular velocity $v_{\rm max}$.
During the MCP, a ring of gas located initially at a radius $R_{\rm i}$
must squeeze up to a final radius $R_{\rm f}$, such as
$v_{\rm max} R_{\rm f}=\sqrt{GM_{\rm i}R_{\rm i}}$ from angular momentum
conservation.
Therefore, the initial radius is given by the simple expression
$R_{\rm i}=0.38 (M_{\rm f}/M_{\rm i})(R_{\rm f}/R_{\rm d})R_{\rm f}$. 
For a normal spiral galaxy like the Milky Way, with a nearly flat 
rotation curve up to the last measured
point, typically at $\sim 30$ kpc, and with $R_{\rm d}\approx 4$ kpc,
the MCP should be able to squeeze the disk at least from $\sim 85$
kpc to $\sim 30$ kpc in radius, provided the mass loss
in outflows is negligible, corresponding to a typical
radial velocity of $3$ km s$^{-1}$ for one Hubble time.
According to \S \ref{sec:core}, this contraction
is an unfeasible process due to the global expansive nature
of magnetic fields.

The level of contraction may be significantly reduced if a
substantial fraction of the mass is lost in winds following early
epoch of star bursts, i.e.~$M_{\rm f}<M_{\rm i}$. Recently, \citet{zha02} has
estimated that less than $4\%$ of the total mass of the Galaxy
has been lost during its lifetime. Even assuming an amount of mass loss
$3$ times larger than this value, the
magnitude of the contraction of the disk is still huge, 
from $72$ kpc to $30$ kpc, again for $R_{\rm d}=4$ kpc.

The MCP represents a serious threat to the suggestion
by \citet{bat02} that, 
since stars do not feel the galactic magnetic field, the
escape of stars formed beyond a certain galactocentric radius could explain
the observed truncation of stellar disks. 
The calculations by Battaner et al.~(2002) are rooted in
a scenario that does not consider the MCP, with the gas disks being
in anomalous rotation from the very beginning. If one tries to include
the MCP in the model, the stellar distribution function
depends on different assumptions that concern the details of
the MCP, such as the radial velocity of the contraction, 
or the relationship between star formation rate,
surface density and magnetic field strength. Consequently, 
the escape radius depends on time.
If the disk was contracted from $\sim 80$ to $30$ kpc in radius,
it could be even more natural to construct a model in which
the stellar surface density decays with $R$ slowly enough, with no stellar
truncation radius. 

Even assuming that stars in the outer disk were not formed
until the gas disk stopped its contraction, which is in contrast
with observations that show old stars in the outer disk (e.g.,
Ferguson \& Johnson 2001), the planar velocity dispersions of the
stars should increase dramatically with $R$ and this is not
observed (e.g., Pfenniger et al.~1994; Vall\'{e}e 1994).

\subsection{The Flaring Problem}
\label{sec:flaring}
In the past years, much of the challenge involved in making
the magnetic alternative viable concerned the excessive flaring
of the disk, which is caused by the vertical magnetic pressure
\citep{cud93}.
In a two-dimensional analysis, \citet{bat95} predict a scale height of
$3$--$3.5$ kpc at a radius of $15$ kpc and of $7$ kpc at
$30$ kpc for the H\,{\sc i} disk of M31.
These values are still too large by a factor of $3$--$5$ if 
comparing with the observed H\,{\sc i} scale height observed in the Galaxy
by \citet{dip91} and \citet{bur92}.

This discrepancy in the scale height is the least of our
worries. An inspection of the radial profile of the magnetic
strength (fig.~3 of Battaner \& Florido 1995) reveals that there is
something deeply wrong in their calculations. 
As shown in \S\ref{sec:oned}, the magnetic
field cannot decrease faster than $1/R$ in order to have a disk with
a faster rotation. 
Therefore, a magnetic field of $5\,\mu$G at $15$ kpc implies a field strength
larger than $2.5\,\mu$G at $30$ kpc. The magnetic field estimated
by \citet{bat95,bat00} decays faster than $1/R$. 

If the dark matter contribution to the rotation curve is null, 
the required magnetic fields in the outer parts of typical spiral galaxies
are so intense ($\sim 5$--$10\,\mu$G) 
that the problem of excessive flaring is unresolved.
The flaring problem is exacerbated when the magnetic pressure of the
small-scale magnetic field is included because it also
contributes to the vertical expansion of the disk. This pressure is observed
to be comparable to the magnetic pressure of the regular field
at least in the optical disk of the Galaxy. On the other hand,
the excessive flaring is also aggravated if the magnetic pitch angle is not
zero. {\it We conclude
that the flaring problem by itself is sufficient to reject the magnetic
alternative.}

\section{Galactic Disks with Slower Rotation: The Cuspy Halo
Problem} 
\label{sec:dwarf}
In this Section, we will discuss the role of magnetic fields 
as a remedy for the cuspy problem of dark halos.
It is interesting to see whether 
the magnetic pressure of the random component,
as well as certain configurations of a 
global-scale magnetic field (not necessarily azimuthal) 
could help to explain the discrepancy between the profiles
of dark halos obtained in cold dark matter (CDM) simulations, 
which becomes very cuspy towards
the center (e.g., Navarro, Frenk \& White 1996, 1997),
and observations of the halos of dwarf galaxies
and LSB, which are better fitted with a core-dominated halo
(see de Blok, Bosma \& McGaugh 2003, and references therein).
One possibility is that galaxies are indeed embedded in cuspy halos
following the profile found by Navarro, Frenk and White (1996; hereafter 
NFW) but the combination of asymmetric drifts and magnetic effects
wipe out completely the inner cusp,
leading to rotation curves that mimic a halo with a core.

The magnetic pressure of the random component gives some support 
to the disk, reducing the rotation speed. Since it is expected to 
be in equipartition with the gas turbulent pressure, corrections
due to the magnetic pressure should be of the order of the asymmetric
drift, $\sim 2$--$3$ km s$^{-1}$ (e.g., de Blok \& Bosma 2002), which
are smaller or comparable to observational uncertainties,
typically $4$--$6$ km s$^{-1}$. This contribution, however, could be larger
in a certain radial portion of the disk if the magnetic pressure decays 
faster than $v_{\rm t}$. Writting the 
Alfv\`en speed associated to the random field 
as $v_{\rm A}^{2}=g(R) v_{\rm t}^{2}$, 
with the turbulent velocity dispersion $v_{\rm t}$ constant with $R$,
and $dg/dR<0$, the rotation curve changes
according to the relation:
\begin{equation}
v_{\phi}^{2}=v_{\rm c}^{2}+g(R)\, \frac{v_{\rm t}^{2}}{2}
\left(\frac{d\ln\rho}{d\ln R} +\frac{d\ln g}{d\ln R}\right).  
\label{eq:drift}
\end{equation}
Following \citet{deb02}, we define $R_{\rm in}$ as the
radius of the innermost sampled point of the 
rotation curve, and assume that at $R>R_{\rm in}$,
$g(R)=(R_{\rm in}/R)^{n}$ for simplicity. Doing so, we are imposing
that at $R_{\rm in}$, $v_{\rm A}\approx v_{\rm t}$.  
The second term of the RHS of Eq.~(\ref{eq:drift}) is of the order
of the asymmetric drift (comparable to observational uncertainties), whereas
the last term is
$-nv_{\rm t}^{2}R_{\rm in}^{n}/2R^{n}$. As the observed rotation curve differs
from the NFW profile along a significant range of galactocentric
radius, say at $R\gg R_{\rm in}\approx 500$ pc, then $0<n\leq 1$, implying that
$nv_{\rm t}^{2}R_{\rm in}^{n}/2R^{n}<v_{\rm t}^{2}/2\sim 50$ km$^{2}$ 
s$^{-2}$ (for $v_{\rm t}=10$ km s$^{-1}$). 
Consequently, this effect   
would be only significant very close to the galactic center where 
$v_{\rm c}\leq 10$ km s$^{-1}$, a region with large observational
uncertainties. Summing up, the contribution from the
total pressure to the circular velocity is less than $5$--$6$ km s$^{-1}$.
This result agrees with the fact that CO, H${\alpha}$ and H\,{\sc i} 
rotation curves reveal the same kinematics
for the central regions of some galaxies (e.g., Bolatto et al.~2002),
which suggests that magnetic fields do not contribute significantly to
the support of the gas disks. In particular, the CO rotation curves 
are expected to trace the motion of
the highest density regions, including molecular clouds, which are
thought to be hardly affected by the galactic magnetic fields. 

In conclusion, magnetic effects 
are not sufficient to solve the cuspy halo problem satisfactorily.

\section{Final Remarks}
\label{sec:final}
It has been recognized for decades that the galactic magnetic field
provides significant vertical support to the interstellar medium.
The role of the magnetic tension in the vertical hydrostatic configuration
was considered by \citet{cox88} and \citet{bou90}, whereas the
role of the tension in the plane of the disk was studied by
\citet{nel88}, \citet{bat92} and \citet{ben00} [see also Piddington 1964].
The idea that large-scale magnetic fields can explain
the rotation curves of spirals without invoking dark matter 
has been revived by \citet{bat00} and Battaner et al.~(2002).

All the previous work aimed at giving bounds to magnetic
effects on gas dynamics are based on
the flaring of the disk by the vertical magnetic gradients. Here we 
propose that the radial analysis provides stringent constraints, 
being free of other complicated assumptions such as the unknown
external pressure.
We have shown that magnetic fields have a net
expansive (or null, but never confining)
effect {\it not only} in the vertical direction but also
in the radial direction. Without dark matter halos, the contribution of the
magnetic field to the rotation curves at the periphery of
spiral galaxies is not expected to be larger than $10$ km s$^{-1}$,
which is insufficient to banish dark halos by magnetic fields.
This result is valid even though an external pressure, intergalactic magnetic
fields or ram pressure are taken into account. 
In addition, the huge contraction of the disk required in the magnetic
alternative would spoil a simple magnetic explanation
for the observed truncation of stellar disks.
Moreover, the problem of excessive flaring in the outer disk by
the magnetic pressure is unresolved and may be aggravated if
the random component of the magnetic field is included, or
if the field is not purely toroidal. By contrast, there is observational
evidence that the dwarf, dark-matter dominated galaxy 
IC 2574 does not possess a dynamically-important galactic-scale
magnetic field. Our first conclusion is that the magnetic alternative
is not viable.

Even though there is no doubt about the existence of dark matter halos
around galaxies, it is desirable to constrain the magnetic effects on
the H\,{\sc i} rotation curves for a correct interpretation of the
observations. 
It is suggested that magnetic fields can contribute to produce wiggles in the
H\,{\sc i} rotation curves, which could be likely associated with
non-axisymmetric waves in the disk, as may be the case of NGC 1560.
However, the magnetic tension is unlikely
to produce substructure in the rotation curve of massive spiral
galaxies above $20$ km s$^{-1}$.
In \citet{san97}, the detailed shape of the rotation curve of this
dwarf galaxy was fitted reasonably well combining a magnetic field 
of strength $\sim 1$ $\mu$G plus an isothermal dark halo.
Finally, we have considered magnetic fields as a possible solution
to the cuspy problem of dark halos. We conclude 
that they are insufficient to explain the discrepancy
between the observations and the predictions of CDM simulations.

\acknowledgements
We thank J.~Cant\'{o} and A.~Hidalgo for discussions, and the anonymous 
referees and the Editor for critical comments. F.J.S.S.~acknowledges 
financial support from CONACYT project 2002-C40366.

\appendix
\section{A. Why an External Pressure is Unimportant to Study the
Radial Configuration}
\label{sec:pressure}

Consider a three-dimensional axisymmetric gas disk with
surface density $\Sigma (R)$ supported mainly by rotation
and embedded in an ambient medium of pressure $P_{\rm ext}$.
The transition between the gas disk and the external medium
occurs at a vertical height $z_{\rm S}(R)$.
From the vertical hydrostatic equilibrium, the radial
gradient of the total (turbulent plus
magnetic) gas pressure $P_{\rm tot}$ at the midplane 
of the disk is given by
\begin{equation}
\frac{dP_{\rm tot}(R,0)}{dR}=\frac{d}{dR}\left[P_{\rm ext}
+\int_{0}^{z_{\rm S}(R)} \rho \,\frac{\partial\Phi}{\partial z}\,dz\right]=
\frac{d}{dR}\left[\int_{0}^{z_{\rm S}(R)} \rho \,\frac{\partial\Phi}
{\partial z}\,dz\right],
\end{equation}
where $\Phi (R,z)$ is the gravitational potential.
The pressure exerted by the ambient medium 
squeezes the disk along $z$, increasing $P_{\rm tot}(R,0)$ but
barely $dP_{\rm tot}(R,0)/dR$,
and thereby making the radial confinement extremely difficult.  
This can be visualized in the following manner.
Ignoring for a moment magnetic fields,
an external pressure will produce some radial confinement if 
it is possible to reach a configuration with
$\partial P/\partial R\big|_{z=0}>0$ just by increasing $P_{\rm ext}$:
\begin{equation}
\frac{d}{dR}\left[\int_{0}^{z_{\rm S}(R)} \rho\,
\frac{\partial \Phi}{\partial z}\,dz\right]>0.
\label{eq:squeeze}
\end{equation}
For a gas density decaying along $z$ with a scale height $h$,
for instance $\rho \propto \exp(-z^{2}/2h^{2})$ (for $|z|< z_{\rm S}$), 
Eq.~(\ref{eq:squeeze}) is satisfied if the scale height 
increases with galactocentric
radius at least as fast as $h\sim z_{\rm S} \sim R^{l}\Sigma^{-1}$,
where $l$ is an exponent that 
varies from $2$--$3$ for a gravitational potential with 
a flat and a Keplerian circular speed, respectively.
However, this requirement cannot be fulfilled in approximately 
isothermal galactic disks with the surface density decreasing outwards,
because the maximum variation of $h$ occurs in the limit of large
external pressure and goes as $h\sim R^{l}$.

There are other physical difficulties in the boundaries of the disks
if they were confined by $P_{\rm ext}$. 
From dimensional grounds, it is expected that the gas
pressure $P(R,0)$, comprising mainly thermal and turbulent contributions as
defined in \S \ref{sec:oned},
should be greater than the external pressure, as a consequence of
turbulent motions and mixing by Rayleigh-Taylor type instabilities in the disk.
The total pressure at $z=0$ is
\begin{equation}
P_{\rm tot}(R,0)=P(R,0)+\frac{B^{2}_{\phi}(R,0)}{8\pi}
>P_{\rm ext}+\frac{B^{2}_{\phi}(R,0)}{8\pi}.
\label{eq:Ptot}
\end{equation}
In the periphery of the disk, pressure confinement requires 
$P_{\rm tot}(R,0)\approx P_{\rm ext}$, thus
Eq.~(\ref{eq:Ptot}) implies $B^{2}_{\phi}(R,0)/8\pi\ll P_{\rm ext}< P(R,0)$, 
i.e.~well below equipartition and, therefore,
they are not sufficient to affect appreciably the rotation curves,
according to our discussion in \S \ref{sec:oned}.

\section{B. Derivation of the Magnetic Terms in the Virial Theorem}
\label{sec:constraints}

Here we discuss the magnetic terms that appears in the virial theorem.
The same derivation can be found in Fiege \& Pudritz (2000).
Consider a distribution of a electrically conducting plasma with
velocity ${\mbox{\boldmath{$v$}}}$ in a 
gravitational field $\Phi(R)$ created by all masses in the system.
The scalar virial theorem 
applied to a finite volume $V$, bounded by a 
surface $S$, can be written as:
\begin{equation}
2{\mathcal{T}}+{\mathcal{U}}+{\mathcal{M}}_{\rm cyl}+{\mathcal{W}}-
\int_{S}\rho\,({\mbox{\boldmath{$v$}}}\cdot
{\mbox{\boldmath{$r$}}})({\mbox{\boldmath{$v$}}}\cdot
d{\mbox{\boldmath{$S$}}})=\frac{1}{2}\frac{d^{2}I}{dt^{2}},
\end{equation}
where ${\mathcal{T}}$ is the kinetic energy of the plasma within $V$, 
${\mathcal{W}}$ the
gravitational term given by ${\mathcal{W}}=-\int_{V} \rho \,
{\mbox{\boldmath{$r$}}}\cdot {\mbox{\boldmath{$\nabla$}}}\Phi \,dV $ and
\begin{equation}
{\mathcal{U}}=3\int_{V}P\,dV-\int_{S} P{\mbox{\boldmath{$r$}}}
\cdot d{\mbox{\boldmath{$S$}}},
\end{equation}
\begin{equation}
{\mathcal{M}}=\frac{1}{8\pi}\int_{V} B^{2}\,dV+
\frac{1}{4\pi}\int_{S} ({\mbox{\boldmath{$r$}}}\cdot
{\mbox{\boldmath{$B$}}})({\mbox{\boldmath{$B$}}}\cdot
d{\mbox{\boldmath{$S$}}})
-\frac{1}{8\pi}\int_{S} B^{2}{\mbox{\boldmath{$r$}}}
\cdot d{\mbox{\boldmath{$S$}}},
\label{eq:redund}
\end{equation}
\begin{equation}
I=\int_V \rho\, r^{2} dV,
\end{equation}
where ${\mbox{\boldmath{$r$}}}$ is a coordinate vector
(e.g., Shu 1992). 

In the case of having cylindrical symmetry
(i.e., $\partial/\partial z=\partial/\partial\phi=0$) and for
fields with no radial component, $B_{R}=0$, a more simple expression for ${\mathcal{M}}_{\rm cyl}$,
which denotes ${\mathcal{M}}$ under cylindrical symmetry,
can be derived when the virial theorem is applied to a finite cylindrical
volume; this is done in the following. 
Since $B_{R}=0$, the second term in the RHS of Eq.~(\ref{eq:redund}) 
contributes only at $z=\pm L/2$ surfaces, where $L$ is the vertical
length of the volume $V$. The component of the field perpendicular
to these surfaces is $B_{z}$, so that
\begin{equation}
\frac{1}{4\pi}\int_{S} ({\mbox{\boldmath{$r$}}}\cdot
{\mbox{\boldmath{$B$}}})({\mbox{\boldmath{$B$}}}\cdot
d{\mbox{\boldmath{$S$}}})=\frac{L}{4\pi}\int_{0}^{R_{S}}2\pi R B_{z}^{2} dR
=\frac{1}{4\pi}\int_{V} B_{z}^{2} dV.
\end{equation}  
The last term of Eq.~(\ref{eq:redund}) comprises an integral along the vertical
surface of the cylinder $I_{1}$ and a contribution of the cap surfaces, 
$I_{2}$, given by 
\begin{equation}
I_{1}=-\frac{1}{4\pi}(B_{\phi,S}^{2}+B_{z,S}^{2})V,
\end{equation}
and
\begin{equation}
I_{2}=-\frac{L}{8\pi}\int_{0}^{R_{S}} 2\pi RB^{2} dR=-\frac{1}{8\pi}\int_{V} B^{2} dV.
\end{equation}
Combining the three contributions the final expression for 
${\mathcal{M}}_{\rm cyl}$ reads as
\begin{equation}
{\mathcal{M}}_{\rm cyl}=\frac{1}{4\pi}\left(\int_{V} B_{z}^{2} \,dV-
\left(B_{z,{\rm S}}^{2}+B_{\phi,{\rm S}}^{2}\right) V\right).
\end{equation}
Similar manipulations allow us to rewrite ${\mathcal{U}}$ as
${\mathcal{U}}=2\int_{V} P\,dV-2P_{S}V$.

For a cold galactic disk, the thermal energy
is negligible compared to the kinetic energy. If, in addition,
there is no radial accretion flows and the disk can be assumed to be 
in a steady state, the virial theorem
is reduced to
\begin{equation}
2{\mathcal{T}}+{\mathcal{M}}_{\rm cyl}+{\mathcal{W}}=0.
\end{equation}
Assuming that galaxies are isolated systems, 
$B_{z,{\rm S}}V$ and $B_{\phi,{\rm S}}V$ go to
zero at large $R$ implying that always ${\mathcal{M}}_{\rm cyl}\geq 0$,
reflecting the impossibility of magnetic confinement
in the {\it radial} 
direction. In fact, for a given distribution of mass, ${\mathcal{W}}$
is fixed and if ${\mathcal{M}}_{\rm cyl}$ is positive,
the total kinetic energy should be smaller than in the unmagnetized
case in order to satisfy the virial constraint.
This does not exclude the possibility that 
some parts of the disk could present a faster rotation (larger
kinetic energy) as long as rotation is slower than that given by
gravity in other parts of the disk.
It might be that we are observing just the part of faster
rotation and the material with a slower rotation lies outside
the last point of H\,{\sc i} detection, if it is fully 
ionized by the metagalactic UV background (see Fig.~1 for a visualization). 
For the case of purely azimuthal fields ${\mathcal{M}}_{\rm cyl}= 0$,
so $2{\mathcal{T}}+{\mathcal{W}}=0$. 
Therefore, given a distribution of mass, all the configurations
in equilibrium must have the same ${\mathcal{T}}$ 
regardless of magnetic fields. Note that ${\mathcal{M}}_{\rm cyl}= 0$
does not imply necessarily that the magnetic configuration is force-free.
In the particular case of a force-free magnetic configuration, it can
be shown that ${\mathcal{M}}_{\rm cyl}= 0$ at any arbitrary volume
and hence magnetic fields are neither expansive nor confining.

\section{C. Constraining the Azimuthal Magnetic Field Strength in
Isolated Galaxies}
Consider an isolated disk galaxy comprising of gas and stars 
with a given total surface density $\Sigma (R)$. 
The gas is moving in pure circular
orbits (i.e.~$v_{z}=v_{R}=0$) and threaded by a purely azimuthal magnetic field.
The distribution of mass (stars$+$gas$+$dark matter)
will create a gravitational potential at
$z=0$, $\Phi_{0}(R)$. The gravitational circular velocity (not the observed 
rotation velocity) is defined as $v_{\rm c}^{2}\equiv Rd\Phi_{0}/dR$. 
We will assume for simplicity that along the
thickness of the H\,{\sc i} disk the rotation curve is constant, i.e.
$\partial v_{\phi}/\partial z|_{R}=0$. The disk is supposed to be rotating
with an anomalous rotation speed $v_{\phi}(R)>v_{\rm c}(R)$ out to the
radius $R_{\rm kep}$. 

We split up the rotational
energy, $T$, in two parts $T=T_{0}+\Delta T$, where $T_{0}$ is the kinetic
energy as if the disk, as it is, were rotating in
force balance with the gravitational potential $\Phi_{0}$
without magnetic fields (keeping all the rest
of the variables of the disk $\Sigma(R)$ and so on unchanged). 
More specifically,
$T$ between two arbitrary radii, say $R_{1}$ and $R_{2}$, and 
within $|z|<h_{1}$, where $h_{1}$ is the characteristic scale height 
of the disk, is given by:
\begin{equation}
T\bigg|_{R_{1}}^{R_{2}}=T_{0}\bigg|_{R_{1}}^{R_{2}}+ 
2 \pi \int_{R_{1}}^{R_{2}}\int_{0}^{h_{1}} 
\!\!\rho\,R\,(v_{\phi}^{2}-v_{\rm c}^{2})\;dz\;dR,
\end{equation}
where $\rho(R,z)$.
According to our discussion in Appendix B, 
${\mathcal{M}}_{\rm cyl}=0$ for pure azimuthal fields, and thus 
the total rotational energy of the
disk must be the same with or without magnetic fields.
Therefore, $\Delta T=0$ between $R=0$ and $\infty$, implying
\begin{equation}
\int_{0}^{R_{\rm kep}}\int_{0}^{h_{1}}\rho R\,(v_{\phi}^{2}-v_{\rm c}^{2})\,dz
\,dR=
-\int_{R_{\rm kep}}^{\infty}\int_{0}^{h_{1}}\rho R\,(v_{\phi}^{2}-v_{\rm c}^{2})\,dz\,dR.
\label{eq:mxm}
\end{equation}
Since we have assumed that $v_{\phi}(R)>v_{\rm c}(R)$ at $R<R_{\rm kep}$, 
this equation implies that $v_{\phi}(R)<v_{\rm c}(R)$ outside the
$R_{\rm kep}$ circle.
Using Eq.~(\ref{eq:vphi}), the LHS of the equation above reads
\begin{equation}
\int_{0}^{R_{\rm kep}}\int_{0}^{h_{1}}\rho R\,(v_{\phi}^{2}-v_{\rm c}^{2})\,dz
\,dR=\frac{1}{8\pi}\int_{0}^{R_{\rm kep}}\int_{0}^{h_{1}}
\frac{d(R^{2}B_{\phi}^{2})}{dR}\,dz\,dR=\frac{\lambda_{B}}{8\pi}
h_{1}R_{\rm kep}^{2}B_{\phi,{\rm kep}}^{2},
\end{equation}
where $B_{\phi, {\rm kep}}$ is the field at $R_{\rm kep}$ and
$\lambda_{B}$ is a dimensionless coefficient less than $1$ defined
by the relation $\int_{0}^{h_{1}} B_{\phi}^2 (R_{\rm kep},z)\,dz=\lambda_{B}
h_{1}B_{\phi}^{2}(R_{\rm kep},0)$. 
Analogously, the RHS of Eq.~(\ref{eq:mxm})
can be written as:
\begin{equation}
\int_{R_{\rm kep}}^{\infty}\int_{0}^{h_{1}}\rho R\,
(v_{\rm c}^{2}-v_{\phi}^{2})\,
dz\,dR=\lambda_{\rho}h_{1}\int_{R_{\rm kep}}^{\infty} 
\rho(R,0)R\,(v_{\rm c}^{2}-v_{\phi}^{2})\,dR,
\end{equation}
where $\lambda_{\rho}$ is a dimensionless parameter less than $1$.
Thus, the value of $B_{\phi,{\rm kep}}$ turns out to be
\begin{equation}
B_{\phi,{\rm kep}}^{2}=  \frac{8\pi
\lambda_{\rho B}} {R_{\rm kep}^{2}}
\int_{R_{\rm kep}}^{\infty} \rho(R,0)R\, 
(v_{\rm c}^{2}(R)-v_{\phi}^{2}(R)) \,dR.
\label{eq:result}
\end{equation}
The factor $\lambda_{\rho B}\equiv\lambda_{\rho}/\lambda_{B}$ 
takes into account possible three-dimensional effects.
For a disk with a constant scale height with radius, one would
expect $\lambda_{\rho}/\lambda_{B}\approx 1$, whereas for a
disk that flares, it could be a little bit larger, depending on
the value adopted for $h_{1}$. As $h_{1}\rightarrow 0$,
$\lambda_{\rho}/\lambda_{B}$ goes to $1$, whereas for a value of order of the
scale height, $\simeq 400$ pc, 
$\lambda_{\rho}/\lambda_{B}\sim 2$ could be considered
as an optimistically high value.

The maximum magnetic effect on the rotation velocity within the
$R_{\rm kep}$ circle occurs when $B_{\phi,{\rm kep}}$ takes
its maximum value; this happens when $v_{\phi}(R)=0$ for $R>R_{\rm kep}$ 
\begin{equation}
B_{\phi,{\rm kep}}^{2}\leq B^{2}_{\rm cr}\equiv \frac{8\pi
\lambda_{\rho B}} {R_{\rm kep}^{2}}
\int_{R_{\rm kep}}^{\infty} \rho(R,0)R\, v_{\rm c}^{2}(R)\,dR.
\end{equation}
$B_{\rm cr}$ is an upper limit to the magnetic field strength at $R_{\rm kep}$.

In the standard dark matter scenario and for galaxies with a flat 
gravitational circular speed, 
$v_{\rm c}(R)=$constant, the maximum relative contribution of the magnetic
field, assuming generously that $B_{\phi}(R)=B_{\rm cr}$ at $R<R_{\rm kep}$
(see \S \ref{sec:isolatedgal}), is:
\begin{equation}
\frac{v_{\phi}^{2}-v_{\rm c}^{2}}{v_{\rm c}^2}\bigg|_{R}=2
\lambda_{\rho B} \left(\frac{R_{\rm g}}{R_{\rm kep}}\right)
\left(1+\frac{R_{\rm g}}{R_{\rm kep}}\right)
\exp\left(\frac{R-R_{\rm kep}}{R_{\rm g}}\right),
\label{eq:twenty}
\end{equation} 
where an exponential law with scalelength $R_{\rm g}$ was adopted for
$\rho(R,0)$ beyond $R_{\rm kep}$. 

In the other extreme case that no contribution to the gravitational
potential comes from a dark component,
$v_{\rm c}^{2}(R)\approx GM/R$ at large $R$, where $M$ is
the mass of the luminous matter in the disk. 
The corresponding maximum relative variation is given by   
\begin{equation}
\frac{v_{\phi}^{2}-v_{\rm c}^{2}}{v_{\rm c}^2}\bigg|_{R}=2
\lambda_{\rho B} \left(\frac{R}{R_{\rm kep}}\right)
\left(\frac{R_{\rm g}}{R_{\rm kep}}\right)
\exp\left(\frac{R-R_{\rm kep}}{R_{\rm g}}\right).
\end{equation}

If the disk is initially, in the lack of magnetic fields,
rotating with velocity $v_{\rm c}$, the
growing magnetic tension during the MCP 
will produce a contraction of the inner
disk and a radial drift of the outer disk.
Let us assume that the disk beyond $R_{\rm kep}$ expands 
in a self-similar fashion and achieves a final density profile
$\rho_{\rm f}(R,0)=(\rho_{0}/\psi^{2})\exp(-R/\psi R_{\rm g})$, with
$\psi >1$ the scale factor. Tthe conservation of angular momentum implies 
that $v_{\phi}(R)=\psi^{-1}v_{\rm c}(R/\psi)$, where $v_{\rm c}(R/\psi)$
is the circular velocity at $R/\psi$.
In the conventional dark halo scenario, $v_{\rm c}=$constant so that
$v_{\phi}(R)=v_{\rm c}/\psi$.
Substituting these relations into Eq.~(\ref{eq:result}), we get
\begin{equation}
B_{\phi,{\rm kep}}^{2}=  
\frac{8\pi(\psi^{2}-1)\lambda_{\rho B}v_{\rm c}^{2}}{\psi^{2} R_{\rm kep}^{2}}
\int_{\psi R_{\rm kep}}^{\infty} \rho_{\rm f}(R,0)\, R\,dR.
\end{equation}
Therefore, the relative effect on the rotation curves starting with
a disk in gravitational force balance is 
\begin{equation}
\frac{v_{\phi}^{2}-v_{\rm c}^{2}}{v_{\rm c}^2}\bigg|_{R}=2
\lambda_{\rho B} \left(\frac{\psi^{2}-1}{\psi^{2}}\right)
\left(\frac{R_{\rm g}}{R_{\rm kep}}\right)
\left(1+\frac{R_{\rm g}}{R_{\rm kep}}\right)
\exp\left(\frac{R-R_{\rm kep}}{R_{\rm g}}\right).
\end{equation} 

Similar manipulations allow us to calculate the same relation but 
assuming that galaxies do not contain massive dark halos. The
corresponding calculation will give 
\begin{equation}
\frac{v_{\phi}^{2}-v_{\rm c}^{2}}{v_{\rm c}^2}\bigg|_{R}=2
\lambda_{\rho B} \left(\frac{\psi-1}{\psi^{2}}\right)
\left(\frac{R}{R_{\rm kep}}\right)
\left(\frac{R_{\rm g}}{R_{\rm kep}}\right)
\exp\left(\frac{R-R_{\rm kep}}{R_{\rm g}}\right).
\end{equation}

\clearpage
\begin{figure}
\plotone{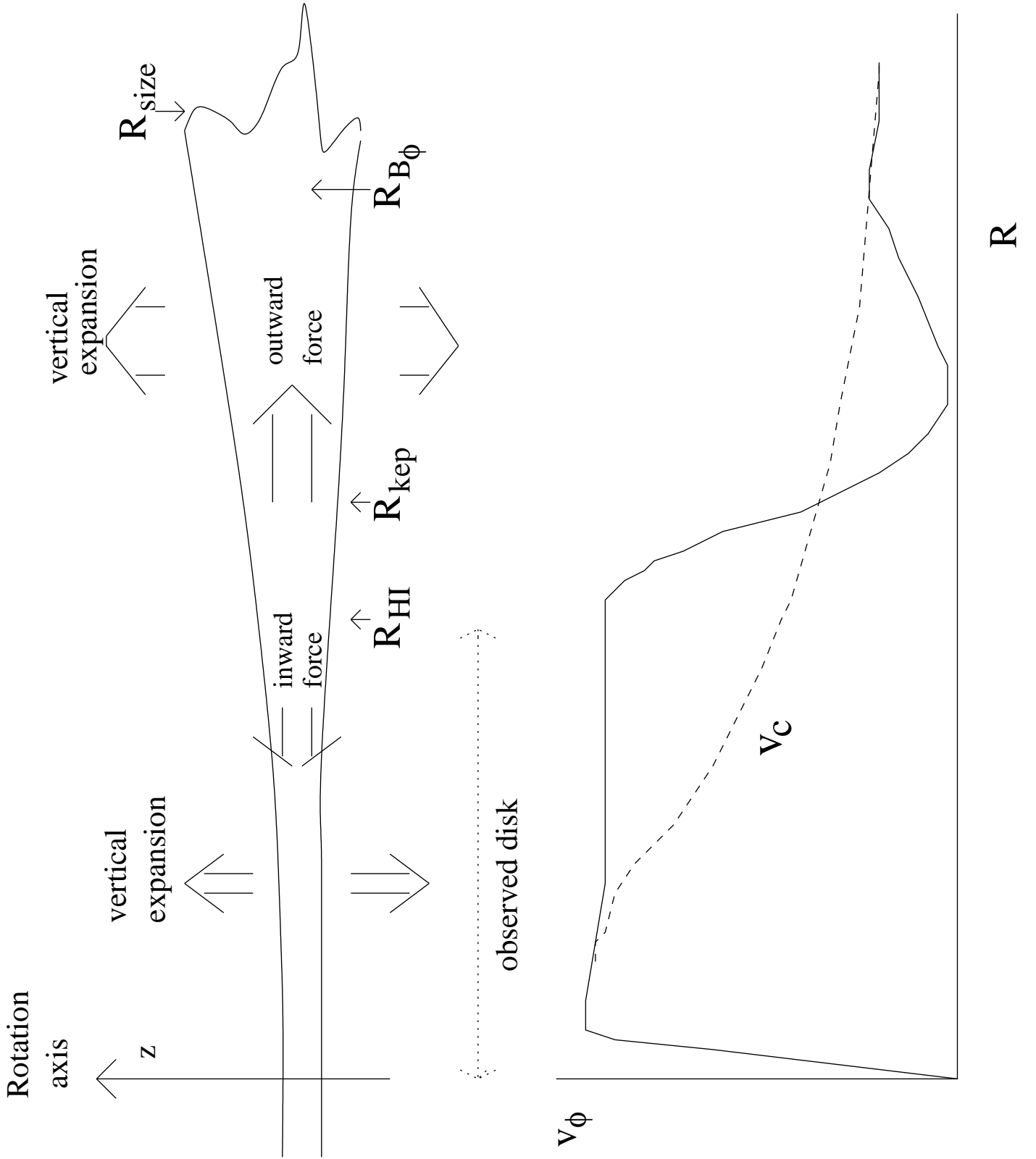}
\caption{Schematic diagram showing the effect of the magnetic
forces in a vertical cross-section of the gas disk in the
magnetic alternative (panel {\it a}). $R_{\rm size}$ is the
radius of the gas disk. Beyond $R_{\rm HI}$, the outermost
radius with H\,{\sc i} detection, the gas could be fully ionized.
The big arrows indicate the net effect of the magnetic fields onto
the gas disk. Beyond $R_{\rm kep}$, the magnetic forces point 
outwards in all directions.
The expected rotation velocity versus radius is 
shown in panel {\it b} (solid line), together with
the gravitational circular rotation speed given by baryonic matter
(dashed line). In both panels $R$ has the same scale.}
\label{sketch}
\end{figure}


\begin{thebibliography}{}
\bibitem[Battaner et al.(1992)]{bat92}
Battaner, E., Garrido, J. L., Membrado, M., \& Florido, E.
1992, \nat, 360, 652
\bibitem[Battaner \& Florido(1995)]{bat95}
Battaner, E., \& Florido, E. 1995, \mnras, 277, 1129
\bibitem[Battaner \& Florido(2000)]{bat00}
Battaner, E., \& Florido, E. 2000, Fund.~Cosmic Phys., 21, 1
\bibitem[Battaner, Florido, Jim\'{e}nez-Vicente(2002)]{bat02}
Battaner, E., Florido, E., \& Jim\'{e}nez-Vicente, J.
2002, \aap, 388, 213
\bibitem[Beck et al.(1996)]{bec96}
Beck, R., Brandenburg, A., Moss, D., Shukurov, A., \& Sokoloff, D.
1996, \araa, 34, 155
\bibitem[Beck(2002)]{bec02}
Beck, R. 2002, in From Observations to Self-Consistent
Modelling of the ISM in Galaxies, ed.~M.~de Avillez, \& D. Breitschwerdt,
\apss, (astro-ph/0212288), in press
\bibitem[Beck(2003)]{bec03}
Beck, R. 2003, in How does the Galaxy work?, ed.~E.~J.~Alfaro,
E.~P\'erez, \& J.~Franco (Kluwer, Dordrecht), (astro-ph/0310287), in press  
\bibitem[Benjamin(2000)]{ben00}
Benjamin, R. A. 2000, RMxAC, 9, 256
\bibitem[Binney \& Tremaine(1987)]{bin87}
Binney, J., \& Tremaine, S. 1987, Galactic Dynamics (Princeton
Univ.~Press: Princeton)
\bibitem[Bland-Hawthorn, Freeman \& Quinn(1997)]{bla97}
Bland-Hawthorn, J., Freeman, K. C., \& Quinn, P. J. 1997, \apj, 490, 143
\bibitem[Bolatto et al.(2002)]{bol02}
Bolatto, A. D., Simon, J. D., Leroy, A., \& Blitz, L. 2002, \apj, 565, 238
\bibitem[Bosma(1978)]{bos78}
Bosma, A. 1978, PhD Thesis, University of Groningen
\bibitem[Boulares \& Cox(1990)]{bou90}
Boulares, A., \& Cox, D. P. 1990, \apj, 365, 544
\bibitem[Bureau \& Carignan(2002)]{bur02}
Bureau, M., \& Carignan, C. 2002, \aj, 123, 1316
\bibitem[Burton(1992)]{bur92}
Burton, W. B. 1992, in The Galactic Interstellar Medium, Saas-Fee
Advanced Course 21, ed. D. Pfenniger \& P. Bartholdi (Berlin: Springer), 1
\bibitem[Carignan, Beaulieu \& Freeman(1990)]{car90}
Carignan, C., Beaulieu, S., \& Freeman, K. C. 1990, \aj, 99, 178 
\bibitem[Castro-Rodr\'{\i}guez et al.(2002)]{cas02}
Castro-Rodr\'{\i}guez, N., L\'{o}pez-Corredoira, M., S\'anchez-Saavedra, M. L.,
\& Battaner, E. 2002, \aap, 391, 519
\bibitem[Chyzy et al.(2000)]{chy00}
Chyzy, K. T., Beck, R., Kohle, S., Klein, U., \& Urbanik, M.
2000, \aap, 355, 128
\bibitem[Cox(1988)]{cox88}
Cox, D. P. 1988, in Supernova Remnants and the Interstellar Medium,
ed. R.S.~Roger, T.L.~Landecker (Cambridge: Cambridge University Press), 73
\bibitem[Cuddeford \& Binney(1993)]{cud93}
Cuddeford, P., \& Binney, J. 1993, \nat, 365, 20
\bibitem[de Blok \& Bosma(2002)]{deb02}
de Blok, W. J. G., \& Bosma, A. 2002, \aap, 385, 816
\bibitem[de Blok, Bosma \& McGaugh(2003)]{deb03}
de Blok, W. J. G., Bosma, A., \& McGaugh, S. 2003, \mnras, 340, 657
\bibitem[Diplas \& Savage(1991)]{dip91}
Diplas, A., \& Savage, B. D. 1991, \apj, 377, 126
\bibitem[Ferguson \& Johnson(2001)]{fer01}
Ferguson, A. M. N., \& Johnson, R. A. 2001, \apj, 559, L13
\bibitem[Fiege \& Pudritz(2000)]{fie00}
Fiege, J. D., \& Pudritz, R. E. 2000, \mnras, 311, 85
\bibitem[Freeman(1997)]{fre97}
Freeman, K. C. 1997, in Dark Matter and Visible Matter in Galaxies,
ASP Conference Series, ed.~M.~Persic, P.~Salucci, 242 
\bibitem[Howard \& Kulsrud(1997)]{how97}
Howard, A. M., \& Kulsrud, R. M. 1997, \apj, 483, 648
\bibitem[Jokipii \& Levy(1993)]{jok93}
Jokipii, J. R., \& Levy, E. H. 1993, \nat, 365, 19
\bibitem[Katz(1994)]{kat94}
Katz, J. I. 1994, \apss, 213, 155
\bibitem[Kronberg et al.(2001)]{kro01}
Kronberg, P. P., Dufton, Q. W., Li, H., \& Colgate, S. A. 2001, \apj,
560, 178
\bibitem[Kulsrud(1999)]{kul99}
Kulsrud, R. M. 1999, \araa, 37, 37
\bibitem[Mestel(1999)]{mes99}
Mestel, L. 1999, Stellar Magnetism (Clarendon Press: Oxford)
\bibitem[Moss \& Shukurov(2001)]{mos01}
Moss, D., \& Shukurov, A. 2001, \aap, 372, 1048
\bibitem[Navarro, Frenk \& White(1996)]{nav96}
Navarro, J. F., Frenk, C. S., \& White, S. D. M. 1996, \apj, 462, 563 (NFW) 
\bibitem[Navarro, Frenk \& White(1997)]{nav97}
Navarro, J. F., Frenk, C. S., \& White, S. D. M. 1997, \apj, 490, 493 
\bibitem[Nelson(1988)]{nel88}
Nelson, A. H. 1988, \mnras, 233, 115
\bibitem[Peratt(1986)]{per86}
Peratt, A. L. 1986, IEEE Trans.~Plasma Sci., vol.~PS-14, no.~6, 763
\bibitem[Persic \& Salucci(1993)]{per93}
Persic, M., \& Salucci, P. 1993, \mnras, 261, L21
\bibitem[Pfenniger, Combes, \& Martinet(1994)]{pfe94}
Pfenniger, D., Combes, F., \& Martinet, L. 1994, \aap, 285, 79
\bibitem[Piddington(1964)]{pid64}
Piddington, J. H. 1964, \mnras, 128, 345
\bibitem[Poezd, Shukurov \& Sokoloff(1993)]{poe93}
Poezd, A., Shukurov, A., \& Sokoloff, D. 1993, \mnras, 264, 285
\bibitem[Reyes-Ruiz \& Stepinski(1996)]{rey96}
Reyes-Ruiz, M., \& Stepinski, T. F. 1996, \apj, 459, 653
\bibitem[Roberts(1967)]{rob76}
Roberts, P. H. 1967, An Introduction to Magnetohydrodynamics 
(Longmans, London)
\bibitem[R\"{o}gnvaldsson(1999)]{rog99}
R\"{o}gnvaldsson, \"{O}. E. 1999, PhD Thesis, University of Copenhaguen
\bibitem[Ruzmaikin, Shukurov \& Sokoloff(1988)]{ruz88}
Ruzmaikin, A. A., Shukurov, A. M., \& Sokoloff, D. D. 1988, 
Magnetic Fields of Galaxies (Kluwer, Dordrecht)
\bibitem[Ryu, Kang \& Biermann(1998)]{ryu98}
Ryu, D., Kang, H., \& Biermann, P. L. 1998, \aap, 335, 19
\bibitem[S\'{a}nchez-Salcedo(1996a)]{sn96a}
S\'{a}nchez-Salcedo, F. J. 1996a, PhD Thesis, Universidad de Granada
\bibitem[S\'{a}nchez-Salcedo(1996b)]{sn96b}
S\'{a}nchez-Salcedo, F. J. 1996b, \apj, 467, L21
\bibitem[S\'{a}nchez-Salcedo(1997)]{san97}
S\'{a}nchez-Salcedo, F. J. 1997, \mnras, 289, 863
\bibitem[Sellwood \& Balbus(1999)]{sel99}
Sellwood, J. A., \& Balbus, S. A. 1999, \apj, 511, 660
\bibitem[Shafranov(1966)]{sha66}
Shafranov, V. D. 1966, in Reviews of Plasma Physics, vol.~I,
ed.~M.A. Leontovich (Consultants Bureau, New York)
\bibitem[Shu(1992)]{shu92}
Shu, F. H. 1992, The Physics of Astrophysics: Gas Dynamics, vol.~II 
(Sausalito: University Science Books) 
\bibitem[Vall\'{e}e(1994)]{val94}
Vall\'{e}e, J. P. 1994, \apj, 437, 179
\bibitem[van Albada et al.(1985)]{van85}
van Albada, T. S., Bahcall, J. N., Begeman, K., \& Sancisi, R. 1985, \apj, 
295, 305
\bibitem[van den Bosch, Burkert \& Swaters(2001)]{van01}
van den Bosch, F. C., Burkert, A., \& Swaters, R. A. 2001, \mnras, 326, 1205
\bibitem[Walter \& Brinks(1999)]{wal99}
Walter, F., \& Brinks, E. 1999, \aj, 118, 273
\bibitem[Widrow(2002)]{wid02}
Widrow, L. M. 2002, Reviews of Modern Physics, 74, 775 
\bibitem[Zhao(2002)]{zha02}
Zhao, H. S. 2002, \mnras, 336, 159

\end{thebibliography}
\end{document}